\documentclass[aps,preprint,tightenlines,showkeys,nofootinbib,superscriptaddress]{revtex4}

\usepackage{amsmath,amssymb,graphicx}

\newcommand{\eftnopi}{\mbox{EFT($\not \! \pi$)}}
\newcommand{\ND}{N^\dagger}
\newcommand{\CSing}{{\cal C}_0^{(^1 \! S_0)}}
\newcommand{\CTrip}{{\cal C}_0^{(^3 \! S_1)}}
\newcommand{\PSing}{P_a^{(^1 \! S_0)}}
\newcommand{\PTrip}{P_i^{(^3 \! S_1)}}
\newcommand{\aSing}{a^{(^1 \! S_0)}}
\newcommand{\aTrip}{a^{(^3 \! S_1)}}
\newcommand{\rSing}{r^{(^1 \! S_0)}}
\newcommand{\rTrip}{r^{(^3 \! S_1)}}
\newcommand{\VS}{\vec{\sigma}}
\newcommand{\VT}{\vec{\tau}}
\newcommand{\LRD}{\stackrel{\leftrightarrow}{D}}
\newcommand{\aR}{{\cal C}^{(^3 \! S_1-^1 \! P_1)}}
\newcommand{\bR}{{\cal C}^{(^1 \! S_0-^3 \! P_0)}_{(\Delta I=0)}}
\newcommand{\cR}{{\cal C}^{(^1 \! S_0-^3 \! P_0)}_{(\Delta I=1)}}
\newcommand{\dR}{{\cal C}^{(^1 \! S_0-^3 \! P_0)}_{(\Delta I=2)}}
\newcommand{\eR}{{\cal C}^{(^3 \! S_1-^3 \! P_1)}}
\newcommand{\DSing}{\Delta^{(^1\!S_0)}}
\newcommand{\DTrip}{\Delta^{(^3\!S_1)}}
\newcommand{\gSing}{g^{(^1\!S_0)}}
\newcommand{\gTrip}{g^{(^3\!S_1)}}
\newcommand{\daR}{g^{(^3 \! S_1-^1 \! P_1)}}
\newcommand{\dbR}{g^{(^1 \! S_0-^3 \! P_0)}_{(\Delta I=0)}}
\newcommand{\dcR}{g^{(^1 \! S_0-^3 \! P_0)}_{(\Delta I=1)}}
\newcommand{\ddR}{g^{(^1 \! S_0-^3 \! P_0)}_{(\Delta I=2)}}
\newcommand{\deR}{g^{(^3 \! S_1-^3 \! P_1)}}
\newcommand{\bs}[1]{\boldsymbol{#1}}
\newcommand{\dpol}{\bs{\epsilon}_d^*}
\newcommand{\gpol}{\bs{\epsilon}_\gamma^*}

\begin{document}

\title{Two parity violating asymmetries from $n p \rightarrow d \gamma$ in pionless effective field theories}

\author{Matthias R. Schindler}
\email{schindle@ohio.edu}
\affiliation{Department of Physics and Astronomy, Ohio University, Athens, OH 45701}
\author{Roxanne P. Springer}
\email{rps@phy.duke.edu}
\affiliation{Department of Physics, Box 90305, Duke University, Durham, NC, 27708}
\date{July 30, 2009}

\begin{abstract}

We consider parity-violating observables from the processes $\vec n p\rightarrow d \gamma$ and $ np \rightarrow d \overset {\circlearrowleft}{\gamma}$. We perform calculations using pionless effective field theory both
with and without explicit dibaryon fields. After combining these results with ones
we have already obtained on parity-violating asymmetries in $\vec NN$
scattering, experimental input would in principle allow the extraction of all
five parameters occurring at leading order in the parity-violating Lagrangian.

\end{abstract}

\pacs{}

\keywords{Parity violation, effective field theory}

\maketitle

\section{Introduction}\label{Sec:Intro}

Low energy hadronic parity violation is of current interest both because of
recent theoretical developments and because of the many and varied experiments
underway or proposed to study the problem
\cite{JLAB,HIGS,Markoff:2005dm,Stiliaris:2005hh,Lauss:2006es,Bass:2009fs}.
In particular,
parity violation in the two nucleon sector
remains an open problem. In a previous paper \cite{Phillips:2008hn} we
addressed the parity violating asymmetries in $\vec N N$
scattering. Here we consider processes involving photons, in
particular polarized and unpolarized radiative neutron capture on
protons.  Experiments to measure asymmetries in $n p
\rightarrow d \gamma$ are difficult, but are important enough that the
experimental community continues to push for better limits.  With new
results expected in the next several years, it is timely to revisit
the theoretical problem.

The parity-violating (PV) component of hadronic interactions is caused by weak
interactions of quarks contained in the hadrons.  Because of the relative
strength of the weak interaction, its manifestation is highly suppressed
compared to the strong interaction. Therefore, we consider observables that
would be zero without the presence of parity-violating effects.
The reaction $np\rightarrow d\gamma$, with
suitable polarizations of the neutron or photon, allows access to two different
asymmetries: the photon asymmetry $A_\gamma$ in polarized neutron capture, $\vec
n p\rightarrow d \gamma$, and the circular photon polarization $P_\gamma$ in unpolarized
capture, $np \rightarrow d \overset {\circlearrowleft}{\gamma}$. In fact,
due to the difficulty of measuring final state photon circular polarization, the inverse
reaction of deuteron photo-disintegration $\overset {\circlearrowleft}{\gamma} d
\rightarrow n p$ may be more experimentally feasible. 
The asymmetry from this inverse reaction is equal to $P_\gamma$ for exactly reversed kinematics.

Weak interactions are well understood in the context of the standard model,
but we are interested here in weak manifestations in hadrons at energies where
QCD is not perturbative. Therefore, we turn to effective field
theories (EFTs) that allow for a perturbative treatment 
in quantities other than the strong coupling constant.
While it is true that investigating hadronic parity violation necessarily
involves complications from nonperturbative QCD, this fact can be used
as an opportunity for probing nonperturbative QCD phenomena in hadrons.
With nucleons, photons, and the deuteron
as physical degrees of freedom we form the set of leading-order operators that obey the symmetries of QCD, but allow
for parity violation.  Because the processes we are interested in
occur at energies well below where the pion is dynamical, we use
an EFT in which the pion is integrated out, \eftnopi, and its physics
is encoded in low energy constants (LECs).

Traditionally, hadronic parity violation has been studied using either 
the so-called Danilov amplitudes \cite{Danilov} or one-boson-exchange models 
\cite{Michel:1964zz,Desplanques:1976mt,Desplanques:1979hn}. 
While the one-boson-exchange models, in particular the one of Ref.~\cite{Desplanques:1979hn}, have been the standard for analyzing experiments, some possible inconsistencies have emerged
(see e.g.~Fig.~5 in Ref.~\cite{Haxton:2008ci}). The effective field
theory treatment of parity-violating hadronic interactions as
performed in this paper allows for a systematic and model-independent
study of few-body low-energy phenomena. As can be seen from the Lagrangian in
the following section, the
{\eftnopi} approach is more
closely related to the Danilov amplitudes than the boson-exchange
models.
The use of EFTs to study parity violation goes back more than a decade
(see e.g.~\cite{Kaplan:1992vj,Savage:1998rx,Kaplan:1998xi}), with a
comprehensive formulation of both pionless and pionful theories given
in Ref.~\cite{Zhu:2004vw}. The asymmetry from $\vec n p \rightarrow d
\gamma$ has been calculated previously in an EFT that included pions
\cite{Kaplan:1998xi} and using the dibaryon formalism
\cite{Savage:2000iv}. 

In the next section we reiterate the five independent PV operators that occur
at leading order in \eftnopi.  One contributes to the $\vec n p \rightarrow d \gamma$
process and three to $n p  \rightarrow d \overset {\circlearrowleft}{\gamma}\ $. Other linear combinations are
accessible through asymmetries in
$\vec N N$ scattering. However, the power counting in the $NN$ system
is still an open question for some observables.  It is clear that
the $NN$ scattering lengths  $a$ are anomalously large, and that it is
necessary to resum  a polynomial series in $a$.  It is less clear
how to treat the $NN$ effective range $r$.  For some processes, 
particularly those involving the deuteron, a much
improved description of data is found by by treating $r$ as large, or resumming the series in $r$ (see Ref.~\cite{Beane:2000fi}, motivated in part by results in Ref.~\cite{Phillips:1999hh}).
But that may not be true for all processes.  Therefore, in Sec.~\ref{Sec:Results}, we present
calculations of $A_\gamma$ and $P_\gamma$  both in the non-dibaryon formalism, where
$r$ is considered to be of ``natural'' size and terms involving $r$
are treated as higher order, as well as in the dibaryon formalism,
where $r$ is treated as anomalously large and the resummed nucleon
bubbles are encoded into a dynamical ``dibaryon'' field. Stating our
results in both languages will provide the flexibility of developing a
common language with other calculations done with more nucleons and/or
using the non-dibaryon formalism.\footnote{While writing up this work we became aware of
contemporary results on $P_\gamma$ in the dibaryon formalism from Shin, Ando, and Hyun \cite{Shin:2009hi}.}

Due to the difficulties inherent in measuring parity violation in the $NN$ system,
information on $ A_\gamma$ and $P_\gamma$ is sparse. 
Currently the asymmetry $A_\gamma$ from $\vec n p
\rightarrow d \gamma$ is consistent with
zero \cite{Cavaignac:1977uk,Alberi:1988fd}, but an ongoing experimental effort is
expected to improve the current value \cite{Lauss:2006es,Seo}.
Results for the photon polarization $P_\gamma$ from $np\rightarrow d \overset
{\circlearrowleft}{\gamma} $ are also consistent with
zero \cite{Knyaz'kov:1984zz}. As
discussed above, the asymmetry from the inverse reaction is equal
to $P_\gamma$ (for suitable kinematics) and, while
current results are again consistent with zero 
\cite{Alberi:1988fd,Earle:1988fc}, there has been recent interest in
performing this measurement \cite{JLAB,HIGS}.

\section{Lagrangians}\label{Sec:Lag}

In \eftnopi\ nucleons interact through contact interactions.
The leading order operators contain the minimum number of
necessary derivatives. The parity-conserving (PC) Lagrangian is given by
\begin{align}\label{Lag:PC}
\mathcal{L}_{PC} & =  \ND(i D_0 + \frac{\vec{D}^2}{2M})N +\frac{e}{2M}\ND (\kappa_0+\tau_3\kappa_1)\,\bs{\sigma}\cdot\bs{B} N \notag\\
&-\CSing (N^T \PSing N)^\dagger (N^T \PSing N) -\CTrip (N^T \PTrip N)^\dagger (N^T \PTrip N ) + \ldots,
\end{align}
with the normalized projection operators \cite{Kaplan:1998sz}
\begin{equation}
\PSing=\frac{1}{\sqrt{8}} \tau_2 \tau_a \sigma_2 \ , \quad 
\PTrip=\frac{1}{\sqrt{8}} \tau_2 \sigma_2 \sigma_i  \ .
\end{equation}
The $\sigma_i$ and $\tau_a$ are SU(2) Pauli matrices in spin and isospin space, respectively, $D_\mu N$ is the nucleon covariant derivative,
\begin{equation}
D_\mu N = \partial_\mu N +ie\frac{1+\tau_3}{2}A_\mu N,
\end{equation}
and $\kappa_0$ and $\kappa_1$ are the isoscalar and isovector nucleon
magnetic moments.  Eq.~(\ref{Lag:PC}) shows only the leading-order
interaction terms.  In the power counting for \eftnopi\ contributions
from other terms are suppressed by powers of $Q/m_\pi$, where $Q\sim
p\sim 1/a $; $p$ is the relative $NN$ momentum
\cite{vanKolck:1997,Kaplan:1998tg,Gegelia:1998gn} and $a$ is
the scattering length in either the $^1S_0$ or $^3S_1$ channel. Using
the power-divergence subtraction (PDS) scheme
\cite{Kaplan:1998tg,Kaplan:1998we} to renormalize loop diagrams the
low-energy constants (LECs) are given by
\begin{align}
\CSing &= \frac{4\pi}{M}\frac{1}{\frac{1}{\aSing}-\mu}\ , \label{Lag:CSing}\\
\CTrip &= \frac{4\pi}{M}\frac{1}{\frac{1}{\aTrip}-\mu}\ . \label{Lag:CTrip}
\end{align}
Here, $\aSing$/$\aTrip$ are the scattering lengths in the $^1S_0$ and $^3S_1$ channel, respectively, and $\mu$ is the subtraction point.

Since the following calculations contain a deuteron in the final state one has to choose an interpolating field for the deuteron. We follow Ref.~\cite{Kaplan:1998sz} and use $\mathcal{D}_i=\ND \PTrip N$. 

For the leading-order PV Lagrangian we use the partial wave notation of Ref.~\cite{Phillips:2008hn}:
\begin{align}\label{Lag:PV}
\mathcal{L}_{PV}=  -  & \left[ \aR \left(N^T\sigma_2 \ \VS \tau_2 N \right)^\dagger 
\cdot  \left(N^T \sigma_2  \tau_2 i\LRD N\right) \right. \notag\\
& +\bR \left(N^T\sigma_2 \tau_2 \VT N\right)^\dagger  
\left(N^T\sigma_2 \ \VS \cdot \tau_2 \VT i\LRD  N\right) \notag\\
& +\cR \ \epsilon^{3ab} \left(N^T\sigma_2 \tau_2 \tau^a N\right)^\dagger 
\left(N^T \sigma_2  \ \VS\cdot \tau_2 \tau^b \LRD N\right) \notag\\
& +\dR \ \mathcal{I}^{ab} \left(N^T\sigma_2 \tau_2 \tau^a N\right)^\dagger 
\left(N^T \sigma_2 \ \VS\cdot \tau_2 \tau^b i \LRD N\right) \notag\\
& +\left. \eR \ \epsilon^{ijk} \left(N^T\sigma_2 \sigma^i \tau_2 N\right)^\dagger 
\left(N^T \sigma_2 \sigma^k \tau_2 \tau_3 \LRD{}^{\!j} N\right) \right] + h.c.,
\end{align}
where $a\, \mathcal{O}\LRD b = a\,\mathcal{O}\vec D b - (\vec D a)\mathcal{O} b$ with $\mathcal{O}$ some spin-isospin-operator, and 
$$ \mathcal{I}=
\begin{pmatrix} 
1 & 0 & 0 \\
0 & 1 & 0\\
0 & 0 & -2
\end{pmatrix}. $$
As shown in Ref.~\cite{Phillips:2008hn}, this form of the Lagrangian is equivalent to the one given in Ref.~\cite{Girlanda:2008ts}.
(Note, however, the different placement of the gauged derivative from
the ungauged derivatives in the operators given in 
Ref.~\cite{Phillips:2008hn}.  While the
ungauged derivative is unaffected by isospin Pauli matrices, it is
important to maintain consistent placement for the gauged derivative.)

The terms in Eqs.~(\ref{Lag:PC}) and (\ref{Lag:PV}) are considered to be of leading order if we assume that the effective ranges $r$ in the $^1S_0$ and $^3S_1$ channel are ``natural'', i.e.~terms like $r/a$ and $rp$ are numerically suppressed. In an alternative power counting the effective ranges are considered large and have to be resummed to all orders. This is most conveniently achieved by use of dynamical dibaryon fields \cite{Kaplan:1996nv,Bedaque:1999vb,Beane:2000fi}. The PC  dibaryon Lagrangian is given by \cite{Beane:2000fi}
\begin{align}\label{Lag:dbPC}
\mathcal{L}_{PC}^{d} = & \ND(iD_0 +\frac{\vec{D}^2}{2M})N 
-t_i^\dagger \left(i\partial_0 +\frac{\vec D^2}{4M} -\DTrip\right) t_i 
-\gTrip \left[t_i^\dagger N^T \PTrip N + \mbox{h.c.}\right]  \notag\\
& - s_a^\dagger \left(i\partial_0 + \frac{\vec D^2}{4M} - \DSing\right) s_a 
-\gSing \left[s_a^\dagger N^T \PSing N + \mbox{h.c.}\right], 
\end{align}
where $t_i$ and $s_a$ are the dibaryon fields in the $^3S_1$ and $^1S_0$ channel, respectively. The couplings can be determined either by integrating out the dibaryon fields or by reproducing the effective range expansion of the $NN$ scattering amplitude. This leads to (suppressing  channel subscripts)
\begin{equation}
g^2=\frac{8\pi}{M^2 r},\quad \Delta=\frac{2}{M r}\left( \frac{1}{a}-\mu\right).
\end{equation}
Note that these procedures only fix the magnitude of $g$ and not its sign. Since the dibaryon-$NN$ couplings are of leading order, insertions of nucleon loops in the dibaryon propagator are not suppressed. This means that the leading-order dibaryon propagator gets dressed by an infinite series of nucleon loop insertions (e.g., \cite{Beane:2000fi}). The dressed propagator is given by
\begin{equation}
S_{d}(E)=\frac{4\pi}{Mg^2}\frac{1}{\mu+\frac{4\pi}{Mg^2}\Delta-\frac{4\pi}{Mg^2}E+i\sqrt{ME}},
\end{equation}
where we have again dropped the partial wave superscripts. For the dibaryon calculations the choice of the deuteron interpolating field is simply the dibaryon field $t_i$ \cite{Beane:2000fi}.

Parts of the PV dibaryon Lagrangian were given in Refs.~\cite{Savage:2000iv,Hyun:2008hp}. The complete Lagrangian is given by
\begin{align}\label{Lag:dbPV}
\mathcal{L}_{PV}^{d}=  -  & \left[ \daR t_i^\dagger \left(N^T \sigma_2  \tau_2 i\LRD_i N\right) \right. \notag\\
& +\dbR s_a^\dagger  
\left(N^T\sigma_2 \ \VS \cdot \tau_2 \tau_a i\LRD  N\right) \notag\\
& +\dcR \ \epsilon^{3ab} \, (s^a)^\dagger 
\left(N^T \sigma_2  \ \VS\cdot \tau_2 \tau^b \LRD N\right) \notag\\
& +\ddR \ \mathcal{I}^{ab} \, (s^a)^\dagger 
\left(N^T \sigma_2 \ \VS\cdot \tau_2 \tau^b i \LRD N\right) \notag\\
& \left. +\deR \ \epsilon^{ijk} \, (t^i)^\dagger 
\left(N^T \sigma_2 \sigma^k \tau_2 \tau_3 \LRD{}^{\!j} N\right) \right] + h.c.\ .
\end{align}
By performing the path integral over the dibaryon fields in the action we can relate the couplings in the two formalisms:
\begin{equation}\label{Lag:PVdbCoup}
g^{(X-Y)}=\sqrt{8}\frac{\Delta^{(X)}}{g^{(X)}}{\cal C}^{(X-Y)} =
\sqrt{\frac{\pi}{r^{(X)}}}\frac{8}{M}\frac{{\cal C}^{(X-Y)}}{{\cal C}_0^{(X)}}.
\end{equation}
For example,
\begin{align}
\dbR & = \sqrt{8}\frac{\DSing}{\gSing}\bR .\notag\\
\end{align}

\section{Results}\label{Sec:Results}

The invariant amplitude for $np\to d\gamma$ can be parameterized as \cite{Kaplan:1998xi}
\begin{align}\label{Res:AmpPara}
\mathcal{M} =& e X N^T \tau_2 \sigma_2 \left[ \bs{\sigma}\cdot\bs{q}\ \dpol\cdot\gpol - \bs{\sigma}\cdot\gpol \ \bs{q}\cdot\dpol\right]N 
+ ie Y \epsilon^{ijk} {\dpol}^i \bs{q}^j {\gpol}^k \left(N^T \tau_2\tau_3 \sigma_2 N\right) \notag\\
& + ie W \epsilon^{ijk} {\dpol}^i {\gpol}^k \left(N^T \tau_2 \sigma_2 \sigma^j N\right) 
+ e V \dpol\cdot\gpol \left(N^T \tau_2 \tau_3\sigma_2 N \right) +\ldots
\end{align}
where the ellipsis stands for terms that are not needed in our
calculation. $\bs{\epsilon_d}$ and $\bs{\epsilon}_\gamma$ are the
polarization vectors of the deuteron and photon, respectively,
$\bs{q}$ is the outgoing photon momentum and $e>0$. The amplitudes $X$
and $Y$ are parity-conserving, while $V$ and $W$ are
parity-violating. For $Y$ and $V$ the initial $NN$ state is in a
relative $^1S_0$ wave, while for $X$ and $W$ it is in a $^3S_1$ wave.
At leading order $Y$ and $W$ contribute to the photon asymmetry in
$\vec n p \rightarrow d\gamma$, while the circular polarization in $n p \rightarrow
d \overset{\circlearrowleft} {\gamma}$ stems from interference between $Y$ and $V$. Below we
discuss the two processes.\footnote{Some of these results have been previously presented by the authors
\cite{Talks}. However, while writing up this work we became aware of a recent result from Shin, Ando, and Hyun
\cite{Shin:2009hi} on $n p \rightarrow d \overset{\circlearrowleft} {\gamma}$.}

\subsection{Photon asymmetry in $\vec n p \to d\gamma$}

The photon asymmetry $A_\gamma$ for $\vec n p \to d\gamma$ at threshold is defined by
\begin{equation}\label{Res:AsymDef}
\frac{1}{\Gamma}\,\frac{d\Gamma}{d\cos \theta}=1+A_\gamma \cos \theta,
\end{equation}
with $\Gamma$ the $np \rightarrow d \gamma$ width and $\theta$ the angle between the neutron polarization and the outgoing photon momentum. The polarization of the neutron leads to interference between the PC amplitudes $X$ and $Y$ and the PV amplitude $W$. At leading order $X=0$ \cite{Kaplan:1998xi} and the asymmetry $A_\gamma$ is given in terms of the amplitudes by \cite{Kaplan:1998xi}
\begin{equation}
A_\gamma=-2\frac{M}{\gamma^2}\,\frac{\mbox{Re}[Y^* W]}{|Y|^2}\ ,
\end{equation}
where $\gamma=\sqrt{MB}$ is the deuteron momentum with $B$ the deuteron binding energy.

\begin{figure}
\includegraphics[width=0.45\textwidth]{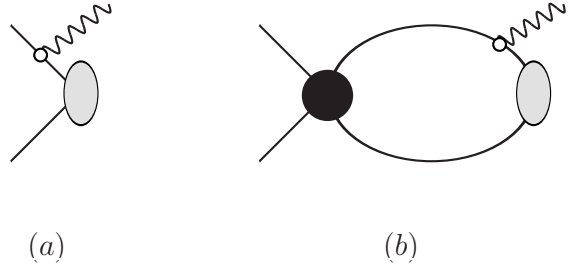}\caption{Leading-order diagrams contributing to the parity-conserving amplitude $Y$. Solid lines denote nucleons, wavy lines denote photons. The large solid circle stands for the resummed $NN$ scattering, the gray oval for the deuteron interpolating field, and the small open circle for a coupling to the nucleon magnetic moment. \label{Fig:Y}}
\end{figure}

The leading-order diagrams contributing to the PC amplitude $Y$ are shown in Fig.~\ref{Fig:Y}, yielding \cite{Kaplan:1998xi}
\begin{equation}\label{Res:Y}
Y=-\frac{2}{M}\sqrt{\frac{\pi}{\gamma^3}}\,\kappa_1 \left(1-\gamma\aSing\right).
\end{equation}

\begin{figure}
\includegraphics[height=0.4\textheight]{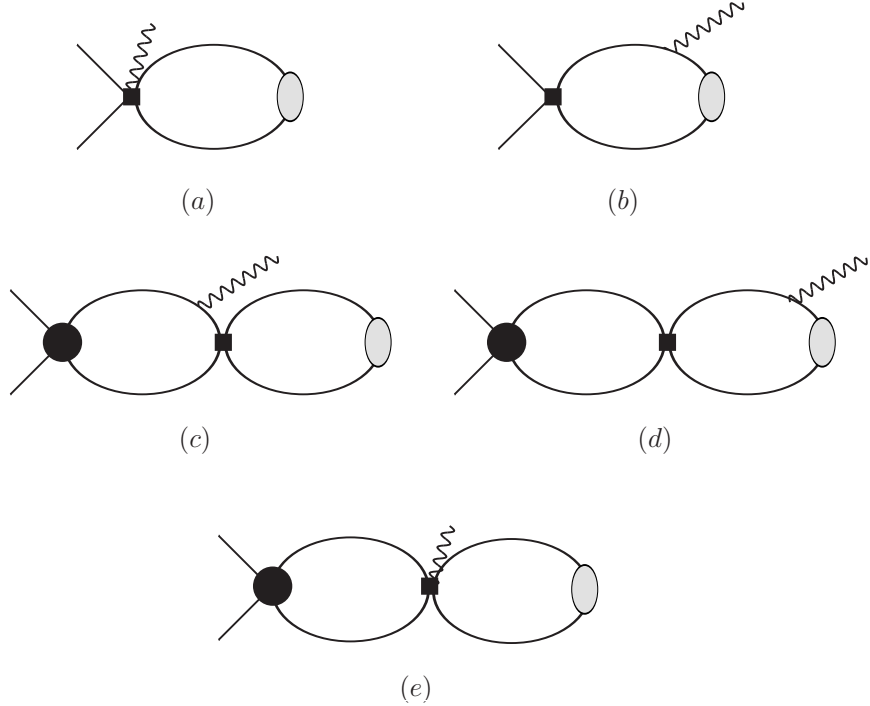}\caption{Leading-order diagrams contributing to the parity-violating amplitudes $W$ and $V$. Solid lines denote nucleons, wavy lines denote photons. The large solid circle stands for the resummed $NN$ scattering, the gray oval for the deuteron interpolating field. The solid
black square is the PV operator.  Photons are minimally coupled . \label{Fig:WV}}
\end{figure}

The diagrams shown in Fig.~\ref{Fig:WV} contribute to the PV amplitude
$W$ for the initial $np$ state in a $^3S_1$ partial wave. Using
$1/\aTrip=\gamma$ (valid at this order),
the result for $W$ is
\begin{equation}\label{Res:W1}
W=\frac{8}{3}\sqrt{\frac{\gamma}{\pi}}\,(\gamma-\mu)\,\eR,
\end{equation}
which, using Eq.~(\ref{Lag:CTrip}) and again $1/\aTrip=\gamma$, can be written as
\begin{equation}\label{Res:W2}
W=\frac{32\pi}{3M}\sqrt{\frac{\gamma}{\pi}}\,\frac{\eR}{\CTrip}.
\end{equation}
This gives the asymmetry
\begin{equation}\label{Res:AsymRes}
A_\gamma=\frac{32}{3}\,\frac{M}{\kappa_1\left(1-\gamma \aSing\right)}\,\frac{\eR}{\CTrip}\ .
\end{equation}
Note the appearance of the ratio $\frac{\eR}{\CTrip}$.
$A_\gamma$ is a physical quantity and must be independent of the subtraction
point $\mu$. $\CTrip$ has the $\mu$ dependence
shown in Eq.~(\ref{Lag:CTrip}), so the $\mu$ dependence of $\eR$ must have the same form  
\begin{equation}
\eR \sim \frac{1}{\frac{1}{\aTrip}-\mu} \ .
\end{equation}
This echos the results discussed in Ref.~\cite{Phillips:2008hn}.

\begin{figure}
\includegraphics[width=0.6\textwidth]{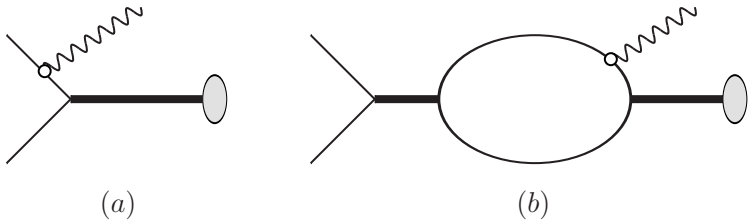}\caption{Leading-order diagrams contributing to the parity-conserving amplitude $Y$ in the dibaryon formalism. Solid lines denote nucleons, thick solid lines denote dressed dibaryons and wavy lines denote photons. The gray oval stands for the deuteron interpolating field, and the small open circle for a coupling to the nucleon magnetic moment. \label{Fig:Ydb}}
\end{figure}

\begin{figure}
\includegraphics[height=0.4\textheight]{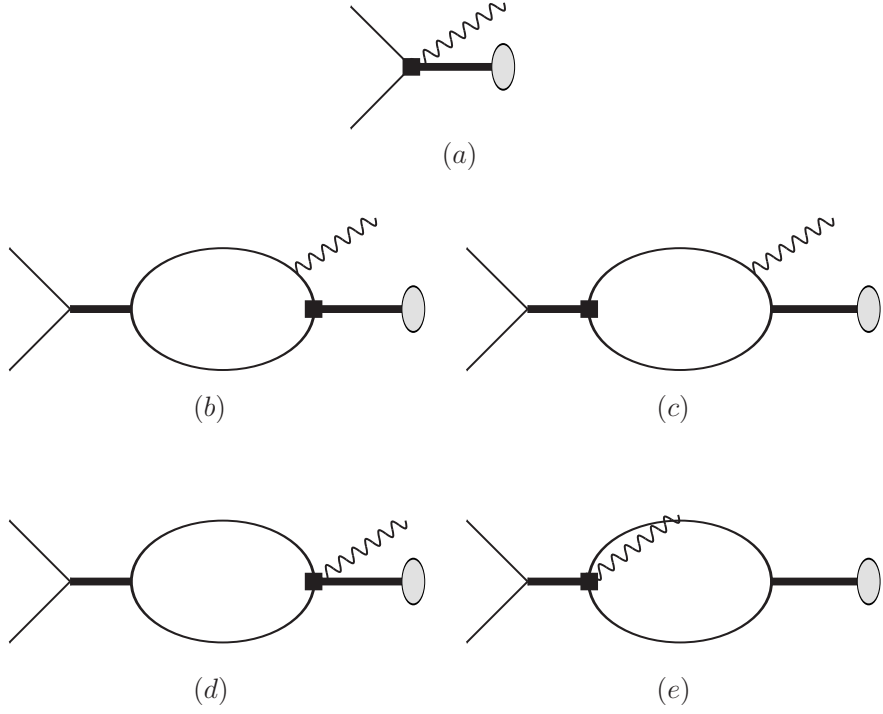}\caption{Leading-order diagrams contributing to the parity-violating amplitudes $W$ and $V$ in the dibaryon formalism. Solid lines denote nucleons, thick solid lines denote dressed dibaryons and wavy lines denote photons. The gray oval stands for the deuteron interpolating field. The solid black box is the PV operator. The photons are minimally coupled.
\label{Fig:WVdb}}
\end{figure}

The diagrams that need to be considered when using the dibaryon Lagrangians of Eqs.~(\ref{Lag:dbPC}) and (\ref{Lag:dbPV}) are shown in Figs.~\ref{Fig:Ydb} and \ref{Fig:WVdb}. In this approach the result for the PC amplitude $Y$ is given by
\begin{equation}\label{Res:Ydb}
Y^d=\frac{2}{M}\sqrt{\frac{\pi}{\gamma^3}}\,\frac{1}{\sqrt{1-\gamma\rTrip}}\, \kappa_1\left(1-\gamma\aSing\right),
\end{equation}
which, when expanded in $\rTrip$, reproduces the result of Eq.~(\ref{Res:Y}) up to a factor of $-1$. Since the amplitude itself is not an observable this sign difference is of no significance and could be absorbed by a field redefinition $t_i\to -t_i$.

For the PV amplitude $W$ we obtain
\begin{equation}\label{Res:Wdb}
W^d=-2\sqrt{\frac{\gamma\rTrip}{1-\gamma\rTrip}}\left(1-\frac{1}{3}\gamma\aTrip\right)\deR.
\end{equation}
Expanding in $\rTrip$ and using Eq.~(\ref{Lag:PVdbCoup}), as well as
$\gamma=1/\aTrip$ at leading order, we reproduce the result of
Eqs.~(\ref{Res:W1}) and (\ref{Res:W2}) up to a factor of $-1$ as
discussed above.\footnote{We note that our values for $W^d$ and $Y^d$ disagree with those in Ref.~\cite{Savage:2000iv} (with $L_1=0$ since it is higher order) by a factor of $-2$ and $2$, respectively.}

The asymmetry is given by
\begin{equation}\label{Res:AsymResdb}
A_\gamma=2M^2\sqrt{\frac{\rTrip}{\pi}}\,\frac{1-\frac{\gamma\aTrip}{3}}{\kappa_1\left( 1-\gamma\aSing \right)}\,\deR,
\end{equation}
which exactly reproduces Eq.~(\ref{Res:AsymRes}) at leading order in $\rTrip$.

\subsection{Circular polarization in $n p \to d \overset{\circlearrowleft}{\gamma}$}

The photon circular polarization in $np\to d \overset{\circlearrowleft}{\gamma}$ is defined by
\begin{equation}\label{Res:PolDef}
P_\gamma=\frac{\sigma_+-\sigma_-}{\sigma_++\sigma_-},
\end{equation}
where $\sigma_{+/-}$ is the total cross section for photons with positive/negative helicity. The polarization again stems from interference between PC and PV amplitudes and, up to the order to which we are working, is given by
\begin{equation} 
P_\gamma=2\frac{M}{\gamma^2} \frac{\mbox{Re}[Y^*V]}{|Y|^2}\ ,
\end{equation}
where we have again used $X=0$ to this order. The expression for $Y$ is already given in Eq.~(\ref{Res:Y}). The amplitude $V$ is calculated from the diagrams in Fig.~\ref{Fig:WV} with the $np$ initial state in a $^1S_0$ partial wave. We find
\begin{equation}\label{Res:V1}
V=4\sqrt{\frac{\gamma}{\pi}}\left[ \left(1-\frac{2}{3}\gamma\aSing\right)(\gamma-\mu)\aR +\frac{\gamma\aSing}{3}\left(\frac{1}{\aSing}-\mu\right)\left(\bR-2\dR\right)\right],
\end{equation}
or, using Eqs.~(\ref{Lag:CSing}) and (\ref{Lag:CTrip}),
\begin{equation}\label{Res:V2}
V=\frac{16\pi}{M}\sqrt{\frac{\gamma}{\pi}}\left[\left(1-\frac{2}{3}\gamma\aSing\right)\frac{\aR}{\CTrip} +\frac{1}{3}\gamma\aSing\frac{\bR-2\dR}{\CSing}\right].
\end{equation}
The photon polarization $P_\gamma$ is given by
\begin{equation}\label{Res:PolRes}
P_\gamma=-16\frac{M}{\kappa_1\left(1-\gamma\aSing\right)}\,\left[ \left(1-\frac{2}{3}\gamma\aSing\right)\frac{\aR}{\CTrip} +\frac{\gamma\aSing}{3}\frac{\bR-2\dR}{\CSing} \right].
\end{equation}

In the dibaryon formalism we obtain
\begin{equation}\label{Res:Vdb}
V^d=-2\sqrt{\frac{\gamma\rTrip}{1-\gamma\rTrip}}\,\left[ \left(1-\frac{2}{3}\gamma\aSing\right)\daR +\frac{\gamma\aSing}{3}\sqrt{\frac{\rSing}{\rTrip}}\left(\dbR-2\ddR\right)\right],
\end{equation}
which leads to the polarization
\begin{align}\label{Res:PolResdb}
P_\gamma = & - 2\sqrt{\frac{\rTrip}{\pi}}\,\frac{M^2}{\kappa_1\left(1-\gamma\aSing\right)} 
\left[ \left(1-\frac{2}{3}\gamma\aSing\right)\daR  \right.\notag\\
&\left.+ \frac{\gamma\aSing}{3}\sqrt{\frac{\rSing}{\rTrip}}\left(\dbR-2\ddR\right)\right].
\end{align}
Using the relations for the PV dibaryon couplings in Eq.~(\ref{Lag:PVdbCoup}) we see that this reproduces the result of Eq.~(\ref{Res:PolRes}).

\section{Conclusion}\label{Sec:Conc}

We obtained results for the photon asymmetry $A_\gamma$ in polarized neutron capture and for the photon polarization $P_\gamma$ in unpolarized capture. We provide results expressed in both the dibaryon and the non-dibaryon coefficient language so that our results can be used in conjunction with additional calculations performed in either operator set. In Eq.~(\ref{Lag:dbPV}) we present the complete leading-order, parity-violating dibaryon Lagrangian required for these calculations. Our results will allow the extraction of two of the PV parameters once the experiments on $n p \rightarrow d \gamma$ become available.

Measurements are available on parity violation in more complicated nuclear (and atomic) systems, but of course these are more difficult for theorists to treat systematically, and to date it is not clear if the experimental results have a consistent theoretical interpretation.  A solid understanding of the two-nucleon sector is likely a necessary prerequisite to understanding the many-nucleon parity violating observables. Fortunately, EFTs allow the extraction of parameters from two nucleon observables that can then be consistently used in calculations on more complicated systems.

To obtain an experimental determination of the five parity violating parameters appearing at leading order, at least five experiments will be required.  The two asymmetries from $np \rightarrow d \gamma$ provide two of them. $\vec n p \rightarrow d \gamma$ provides $\eR$ (or in the dibaryon language, $\deR$) while $ np \rightarrow d \overset {\circlearrowleft}{\gamma}$ yields a linear combination of $\aR$, $\bR$, and $\dR$ (or the corresponding dibaryon  coefficients). Asymmetries from $\vec n n$, $\vec n p$, and $\vec p p$  scattering would
yield three more linear combinations, including a dependence on $\cR$, but
experimental results on $\vec n p$ and $\vec n n$ scattering are
unlikely in the near future. To obtain further constraints on the parameters would require extending the treatment to
few-body systems. From a theoretical perspective a three-body
calculation is the natural extension of the current program, in
particular considering recent experimental interest in the reaction
$\vec n d\rightarrow t \gamma$ \cite{Crawford}.

\begin{acknowledgments}

We thank D.~R.~Phillips for interesting discussions and a careful reading of the manuscript. MRS acknowledges the hospitality of the Lattice and Effective Field Theory group at Duke University. This research was supported by DOE grant DE-FG02-93ER40756 (MRS) and DOE grant DE-FG02-05ER41368 (RPS).

\end{acknowledgments}

\end{document}